\documentclass{appolb}
\usepackage{graphicx}
\usepackage{cite}
\newcommand{\Imag}[1]{\Im {\it m}(#1 )}

\begin{document}
\title{Unquenching weak substructure%
\thanks{Presented at Workshop on Unquenched Hadron Spectroscopy:
Non-Perturbative Models and Methods of QCD vs. Experiment,
Coimbra (Portugal), 1--5 September 2014}%
}
\author{Eef van Beveren
\address{Centro de F\'{\i}sica Computacional,
Departamento de F\'{\i}sica,\\ Universidade de Coimbra,
P-3004-516 Coimbra, Portugal}
\\ [10pt]
George Rupp
\address{Centro de F\'{\i}sica das Interac\c{c}\~{o}es Fundamentais,
Instituto Superior T\'{e}cnico,
Universidade de Lisboa,
P-1049-001 Lisboa, Portugal}
\\ [10pt]
Susana Coito
\address{Institute of Modern Physics, CAS, Lanzhou 730000, China}
}
\maketitle
\begin{abstract}
On assuming that weak substructure has a dynamics which is similar
to quantum chromodynamics but much stronger,
we conclude that unquenching is indispensable
for predictions on the spectrum of weak-substructure resonances.
\end{abstract}
\PACS{
12.40.Yx, 
12.60.Rc, 
13.25.-k  
}

\section{Introduction}

It is well known that com\-po\-si\-te\-ness may be studied
from the appearance of resonance en\-han\-ce\-ments
in the event distributions of scattering and production experiments.
An extensive study on ha\-dro\-nic com\-po\-si\-te\-ness
published by Godfrey and Isgur \cite{PRD32p189}
gave us a good insight into the spectrum of quarkonia
obtained by the scattering of mesons
and by the event distributions of two or more hadrons
produced in production experiments.

Nevertheless, at present our knowledge of hadronic spectra
is limited by the lack of accurate experimental data \cite{ARXIV14114151}.
In particular, any bump in hadronic cross sections
is usually interpreted as a resonance, whereas
resonance structures that are not of a Breit-Wigner-like shape
remain unrecognised in experimental analyses.
Models like that of Ref.~\cite{PRD32p189}
are helpful in order to classify mesonic and baryonic resonances.
However, several details of the spectra remain unsolved.
Here we shall concentrate on mesonic resonances.

Already as early as one decade after the introduction
of the quark model
by Zweig \cite{CERNREPTH401/412} and Gell-Mann \cite{PL8p214},
it was recognised that confinement models alone cannot explain the
event distributions \cite{Cargese75p305}, since mesonic resonances are
observed in scattering and production experiments.
As a consequence, quark confinement and the scattering of hadrons
have to be treated on the same footing.
Scattering is most conveniently described by a scattering amplitude
$T$ as a function of the total invariant mass $\sqrt{s}$.
Mesonic resonances appear as singularities (poles)
of the analytic continuation of $T\left(\sqrt{s}\right)$
into the complex $\sqrt{s}$ plane.
The real and imaginary parts of a pole approximately correspond
to the central mass and the width of a resonance, respectively.
For the scattering of mesons, $T\left(\sqrt{s}\right)$
must contain various disctinct channels, because of the
possible formation of different multi-hadron final states.
Here we shall limit ourselves to final states that contain pairs of
mesons.

In Ref.~\cite{PRD21p772} a model was developed that incorporates
quark confinement in the construction of the scattering amplitude.
The model represents confinement by binding the valence quarks
via a harmonic-oscillator (HO) potential.
Nevertheless, the pole spectrum
of the resulting scattering amplitude
is very different from the HO spectrum.
Moreover, resonances are not represented
by pure HO wave functions
(see e.g.\ Ref.~\cite{ARXIV14117902}),
but rather by several components,
namely for the allowed valence $q\bar{q}$ states with the resonance's
quantum numbers and for the most relevant two-meson channels \cite{PRD21p772}.
In this model it is assumed that mesonic resonances and the free two-meson
states resulting from decay couple to each other via the 
creation or annihilation of new $q\bar{q}$ pairs, with intensity
represented by a parameter $\lambda$. In principle $\lambda$ has to be
adjusted to experiment, but in practice it has been found to be rather
independent of the meson's flavour content \cite{PRD27p1527}.
Furthermore, the internal hadronic dynamics, governed by glue,
appears to be well represented by HO confinement,
with an oscillator frequency $\omega =190$ MeV,
independent of the meson's flavour content.

Let us study, for example, resonances in the vector-charmonium sector.
The quantum numbers of such systems are $c\bar{c}$ for flavour
content and $J^{PC}=1^{--}$ for spin, parity, and $C$-parity.
A $1^{--}$ $c\bar{c}$ system has total quark-antiquark spin $s$=1,
and relative quark-antiquark angular momentum
$\ell$=0 ($S$-wave) or $\ell$=2 ($D$-wave).
A convenient selection of open-charm two-meson states
with $J^{PC}=1^{--}$ may consist of
$D\bar{D}$ with total two-meson spin $S$=0
and relative two-meson angular momentum $L$=1,
$D\bar{D}^{\ast}+\bar{D}D^{\ast}$ ($S$=1/$L$=1),
$D^{\ast}\bar{D}^{\ast}$ ($S$=0/$L$=1, $S$=2/$L$=1, or $S$=2/$L$=3),
$D_{s}\bar{D}_{s}$ ($S$=0/$L$=1),
$D_{s}\bar{D}^{\ast}_{s}+\bar{D}_{s}D^{\ast}_{s}$ ($S$=1/$L$=1),
and $D^{\ast}_{s}\bar{D}^{\ast}_{s}$
($S$=0/$L$=1, $S$=2/$L$=1, or $S$=2/$L$=3).
We then obtain for the description of resonances in the charmonium sector
a coupled system of two quark-antiquark wave functions
describing the probability of finding in the interaction region
a $c\bar{c}$ pair in either of the two possible spatial configurations,
and ten two-meson wave functions representing the probability of finding a
pair of mesons in any of the ten flavour and spatial configurations.

In the model of Ref.~\cite{PRD21p772}, it was assumed that
the two-meson states couple to the $c\bar{c}$ states exclusively
through the creation/annihilation of $u\bar{u}$, $d\bar{d}$,
or $s\bar{s}$ quark-antiquark pairs.
No further final-state interactions within or among the two-meson channels
were considered. Reality is of course somewhat more involved, but as in
any model priorities must be set and further details left for future research.
The used coupling constants were determined in Ref.~\cite{ZPC21p291}
employing an HO approximation.
Thus we find that the $S$=2/$L$=3 two-meson configurations
only couple to the $D$-wave $c\bar{c}$ states, whereas all the other
two-meson configurations couple to both $S$- and $D$-wave $c\bar{c}$ states.
Recently, similar results and some rather interesting consequences were
derived in Refs.~\cite{PRD90p034009,ARXIV14112485}.

The above strategy of describing mesonic resonances via coupled-channel
states was initially baptised the {\it unitarisation} \/scheme,
as it leads to a unitary $S$-matrix instead of just energy levels. However, 
it is nowadays more often called {\it unquenching} the quark model
\cite{ARXIV14117902}, since mesonic resonances are described by coupling
confined (quenched) quark-antiquark states to the meson-meson continuum, just
like in fully unquenched lattice calculations \cite{PRD88p054508}.

Let us now assume for a moment that it was possible to scatter $\bar{D}$
mesons off a source of $D$ mesons. Then one could observe in experiment
the $c\bar{c}$ resonances in the  $D\bar{D}$ scattering cross sections.
However, as it concerns a coupled-channel system, one might also observe
$D_{s}\bar{D}_{s}$ final states, or any of the other flavour and spatial
configurations that couple to $c\bar{c}$.
For this reason, the scattering amplitude for $D\bar{D}$ scattering
is described by a 10$\times$10 complex symmetric matrix
in the model of Ref.~\cite{PRD21p772}. Each one of the 100 elements of that
matrix, when analytically continued to complex invariant mass,
contains the singularities that correspond to the various possible resonances.

For HO confinement we have
an equidistant $J^{PC}=1^{--}$ $c\bar{c}$ spectrum
with spacing $2\omega$, i.e., one $S$-wave ground state
and degenerate pairs of $S$- and $D$-wave excited states.
As a consequence, we expect to find a similar resonance-pole spectrum
for the model. To a certain extent, this is indeed what is
observed in experiment, as we shall discuss below.
In the model \cite{PRD21p772}, the ground state is affected the most by
unquenching and comes out several hundreds of MeV lower than 
the bare mass from confinement only.
Its wave function contains sizable components
in the two-meson channels,
while the $c\bar{c}$-channel is no longer
a pure HO ground state.
On the other hand, the degenerate pairs of $S$- and $D$-waves
show a very peculiar behaviour when the degeneracy is lifted
upon unquenching, with the $S$- and $D$-wave components getting mixed.
Namely, the dominantly $D$-wave mixture almost decouples from
the two-meson channels, whereas the mainly $S$-wave one 
couples much more strongly.
As a result, the mostly $D$-wave mixtures stay near the energy levels of
pure HO confinement, while the others are shifted downwards
about 150--200 MeV.
A further consequence is that the resonance poles
for the mainly $D$-wave mixtures
do not have large imaginary parts and thus are narrow.
This may well explain why they are not easily found in experiment.

\begin{figure}[b]
\centerline{%
\includegraphics[trim = 0mm 2mm 0mm 2mm,clip,width=12.5cm,angle=0]
{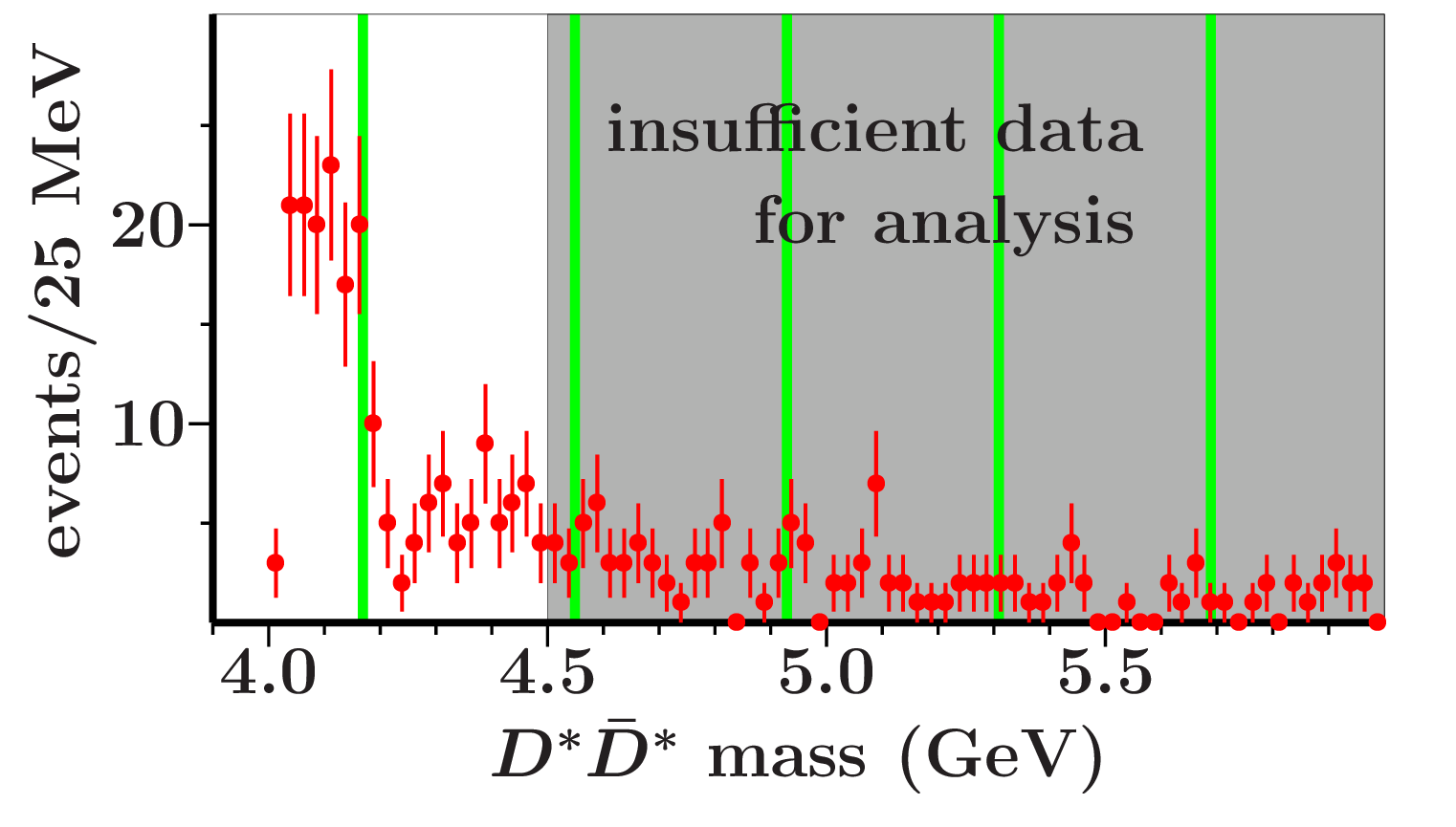}}
\caption{Invariant-mass distribution for $J^{PC}=1^{--}$
$D^{\ast}\bar{D}^{\ast}$ states published by the BaBar Collaboration
\cite{PRD79p092001}.
The vertical lines indicate the spectrum
for HO confinement in the
$J^{PC}=1^{--}$ $c\bar{c}$ sector.
Resonances in the non-shaded area (3.9--4.5 GeV) are known
for almost four decades.
The data in the shaded area (4.5--6.0 GeV) do not have enough statistics.}
\label{charmonium}
\end{figure}

So the dominantly $D$-wave mixtures
of the $J^{PC}=1^{--}$ $c\bar{c}$ spectrum may serve as an indication
for the bare quark-confinement spectrum.
This represents a unique opportunity,
since $J^{PC}=1^{--}$ are precisely the quantum numbers
for electron-positron annihilation.
Hence, in $e^{-}e^{+}$ scattering experiments
one may find a straightforward feedback on quark confinement.
Consequently, the $J^{PC}=1^{--}$ $c\bar{c}$ spectrum
should form the backbone of meson spectroscopy.
Now, what can the experimental state-of-the-art say about that?
In Fig.~\ref{charmonium} we depict the present situation.
It clearly shows that the study of hadronic resonances
is severely hampered by a lack of accurate data.
The charmonium resonances at 4.03, 4.16, and 4.40 GeV
were first observed almost four decades ago.
The data in the invariant-mass interval 4.5--6.0 GeV
do not have enough statistics for further analysis
(see, however, Ref.~\cite{CNPC35p319}).
This issue can only be solved with much better statistics,
and bin sizes that do not exceed 1.0 MeV
in order to discover the narrow $D$ states.

A further consequence of ha\-dro\-nic com\-po\-si\-te\-ness
is the appearance of non-resonant threshold en\-han\-ce\-ments.
A theoretical model for threshold en\-han\-ce\-ments
in ha\-dro\-nic production amplitudes,
based on quark-antiquark pair creation,
was formulated in Ref.~\cite{AP323p1215}
and further developed in Refs.~\cite{EPL81p61002,EPL84p51002}.
This model shows that one must expect non-resonant en\-han\-ce\-ments
in the amplitudes just above pair-creation thresholds.
In the case of stable hadrons, such enhancements are
accompanied by clear minima right at the thresholds,
as observed in experiment for the process $e^{-}e^{+}\to b\bar{b}$,
measured and analysed by the BaBar Col\-la\-bo\-ra\-tion
\cite{PRL102p012001}.
As also remarked by BaBar in their paper,
the large statistics and the small energy steps of the
scan make it possible to clearly observe the dips
at the opening of the thresholds corresponding
to the $B\bar{B}^{\ast}+\bar{B}B^{\ast}$
and $B^{\ast}\bar{B}^{\ast}$ channels.
However, experimental evidence of this phenomenon is scarce,
since it needs event counts with high statistics and good resolution.
Nevertheless, in some cases signals, albeit often feeble,
can be seen in experimental data for ha\-dro\-nic production
\cite{PRD80p074001}.

In Ref.~\cite{AP323p1215} the generic relation
\begin{equation}
P=\Imag{Z}+TZ
\label{production}
\end{equation}
between two-particle scattering ($T$) and production ($P$) amplitudes
was studied in a microscopic multi-channel model for meson-meson scattering
with coupling to confined quark-antiquark channels.
The amplitude $T$ in expression~(\ref{production}) is supposed
to contain the resonance poles that occur in scattering,
whereas $Z$ is a smooth function of invariant mass.
Threshold enhancements occur in production amplitudes
as a consequence of the shape of $\Imag{Z}$,
which in the ideal case of no further nearby thresholds
rises sharply just above threshold.
For larger invariant masses $\Imag{Z}$ first reaches a maximum
and then falls off exponentially.
As a consequence, production amplitudes show non-resonant yet
resonant-like enhancements just above threshold.
In Fig.~\ref{charmonium} one may observe such an enhancement at 4.66 GeV,
just above the $\Lambda_{c}^{+}\Lambda_{c}^{-}$ threshold
\cite{CNPC35p319},
while the large bump at 4.04 GeV
may well consist of the enhancement
above the $D^{\ast}\bar{D}^{\ast}$ threshold
interfering with a $c\bar{c}$ resonance of modest size.

Besides threshold enhacements, unquenching may also dynamically generate
resonances, i.e., resonance poles in the scattering amplitude
that are not directly related to the confinement spectrum.
The low-lying scalar mesons are the classical example
of this phenomenon \cite{ZPC30p615}.
So enhancements can be due to resonances
that are either directly related to the confinement spectrum
or dynamically generated, but may also correspond to non-resonant threshold
effects.  Consequently, analysing mesonic scattering/production data
is a rather difficult task,
in particular when the spatial quantum numbers
are completely/partly unknown,
which unfortunately is most commonly the case.

\section{Weak substructure}

In Refs.~\cite{ARXIV13047711,ARXIV14114151}
we have indicated the possible existence of substructure
in the weak sector, based on the observation that recurrences 
of the $Z$ boson may exist.
The corresponding data, published in
Refs.~\cite{PLB710p403,PLB716p1,PLB479p101,ARXIV12052907,
EPJC74p3076,ARXIV13053315,PRD89p092007,PLB345p609,CMSPASHIG-13-001},
do not have sufficient statistics to definitely conclude
the existence of weak substructure, except perhaps for a clear dip
at about 115 GeV in diphoton, four-lepton, $\mu\mu$,
and $\tau\tau$ invariant-mass distributions.
The latter structure indicates the possible opening of
a two-particle threshold, probably corresponding to a pseudo-scalar partner
of the $Z$ boson with a mass of about 57.5 GeV.
Further possible recurrences of the $Z$ boson, viz.\ at 210 and 240 GeV
\cite{ARXIV13047711}, certainly need a lot more statistics.

Composite heavy gauge bosons and their spin-zero partners,
the latter with a mass in the range 50--60 GeV,
were considered long ago \cite{PLB135p313}
and studied in numerous works (see e.g.\ Refs.\
\cite{PLB141p455,PRAMANA23p607,PRD36p969,NCA90p49,PRD39p3458,PRL57p3245}).
To date, no experimental evidence of their existence has been reported.
However, if a pseudo-scalar partner
of the $Z$ boson with mass of about 57.5 GeV exists
and, consequently, part of the structure
observed in the mass interval 115--135 GeV
is interpreted as a threshold enhancement,
then it must be possible to verify their existence at LHC,
for example in four-photon events.

More recently the interest in weak substructure has revived
\cite{ARXIV12074387,ARXIV12105462,ARXIV13076400,ARXIV13040255,PRD90p035012}.
Most popular among the proposed models is so-called technicolour  (TC)
\cite{PRD20p2619}, for which one expects QCD-like dynamics
but much stronger.
From the structure of the threshold enhancement above 115 GeV,
we deduced an interaction distance of the order of 0.008 fm
\cite{ARXIV14114151}.
Now, from QCD we have learned that self-interactions
lead to an appreciable contribution to the masses of resonances.
Hence, for yet much stronger dynamics we must expect
that the masses of resonances are basically determined
by the self-interactions and not so much
by the masses and binding forces of the constituents.
This has indeed been recognised in Ref.~\cite{PRD90p035012},
where, in a perturbative fashion, the mass of the TC scalar resonance
is lowered by several hundreds of GeV.
However, as we have argued that already for QCD unquenching
should be incorporated beyond perturbative contributions,
we assume that for weak substructure it is absolutely indispensable to
do so.  This also implies that the corresponding spectrum
will contain dynamically generated resonances as well
and may even be dominated by such poles,
rather than by those stemming from confinement.

\section{Conclusions}

Modelling the dynamics of strong interactions is useful
and certainly a lot of fun.
However, it must be accompanied by
the study of scattering and production \cite{ARXIV150101691}.
Experiment, unfortunately, does not yet provide the necessary statistics
to systematically confront model results with measured cross sections.

\newcommand{\pubprt}[3]{#1, #2 (#3)}
\newcommand{\AP}[1]{{\it Ann.\ Phys.} \/{\bf #1}}
\newcommand{\CNPC}[1]{{\it Chin.\ Phys.} \/{\bf C#1}}
\newcommand{\EPJC}[1]{{\it Eur.\ Phys.\ J.} \/{\bf C#1}}
\newcommand{\EPL}[1]{{\it Europhys.\ Lett.} \/{\bf #1}}
\newcommand{\NCA}[1]{{\it Nuovo Cim.} \/{\bf A#1}}
\newcommand{\PL}[1]{{\it Phys.\ Lett.} \/{\bf #1}}
\newcommand{\PLB}[1]{{\it Phys.\ Lett.} \/{\bf B#1}}
\newcommand{\PRAMANA}[1]{{\it Pramana} \/{\bf #1}}
\newcommand{\PRD}[1]{{\it Phys.\ Rev.} \/{\bf D#1}}
\newcommand{\PRL}[1]{{\it Phys.\ Rev.\ Lett.} \/{\bf #1}}
\newcommand{\ZPC}[1]{{\it Z.\ Phys.} \/{\bf C#1}}

\end{document}